\def\cao{\c c\~ao\ }

\def\1{\'{\i}}
\def\beq{\begin{equation}}
\def\eeq{\end{equation}}
\def\bea{\begin{eqnarray}}
\def\eea{\end{eqnarray}}
\def\bed{\begin{displaymath}}
\def\eed{\end{displaymath}}

\def\R{\mathbb{R}}
\def\Z{\mathbb{Z}}

\documentclass[twocolumn,preprintnumbers,prl,aps]{revtex4}
\usepackage{graphics,graphicx,amssymb,amsmath,bm,dcolumn,url}

\begin{document}

\title{Finite-time rotation number: a fast indicator for  chaotic dynamical
structures}

\author{J. D. Szezech Jr.$^1$, A. B. Schelin$^1$, I. L. Caldas$^1$, S. R. Lopes$^2$, P. J. Morrison$^3$, and 
R. L. Viana$^2$%\footnote{Corresponding author. e-mail: viana@fisica.ufpr.br}
}
\affiliation{1.\  Instituto de F\'{\i}sica, Universidade de S\~ao Paulo, 5315-970, S\~ao Paulo, S\~ao Paulo, Brazil. \\2.\  Departamento de F\1sica, Universidade Federal do Paran\'a, 81531-990, Curitiba, Paran\'a, Brazil. \\3.\  Department of Physics, The University of Texas at Austin, Austin, TX 78712.}

\begin{abstract}
Lagrangian coherent structures are effective barriers, sticky regions, that separate phase space regions of different dynamical behavior.   The usual way to detect such structures is via  finite-time Lyapunov exponents. We show that similar results can be obtained for single-frequency systems  from finite-time rotation numbers, which are much faster to compute. We illustrate our claim by considering examples of continuous and discrete-time dynamical systems of physical interest. 
\end{abstract}

\maketitle
\date{today}

It is noteworthy that chaos is  observed in physically interesting systems,  both in Nature and in
mathematical models. Indeed, chaotic dynamics is now commonplace
in such  diverse disciplines as  celestial mechanics
\cite{celestial}, atomic physics \cite{cristel}, fluid mechanics
\cite{aref-ottino}, plasma physics \cite{plasmahorton}, etc. 
The main means of determining chaos is with Lyapunov exponents, which can be notoriously
difficult to compute in fast and reliable ways, particularly if only
experimental data are available and the governing equations are unknown. 
Hence,  an active area of research is
the  search for  {\it fast indicators} (FIs), i.e., computational diagnostics
that characterize chaos quickly \cite{frouard}.   Here we introduce a new 
FI, the {\it finite time rotation number} (FTRN),  for determining effective transport barriers or sticky regions.

Since Lyapunov exponents measure the exponential rate of divergence
of nearby trajectories, FIs  are usually based on
them, examples being the finite-time Lyapunov exponent (FTLE) \cite{froeschle},
the smaller (SALI) and generalized (GALI) alignment indices  
\cite{skokos}, and the mean exponential growth of nearby orbits
(MEGNO) \cite{cincotta}.  These FIs  not only 
determine the strength of chaos,  but can detect invariant tori and other
issues related to integrability. Alternatively,  there are diagnostics  based 
on frequency decomposition, which  are well-suited for
weakly chaotic motion with any  number of degrees of freedom  \cite{laskar}. 

A relatively new application of FIs is to quantitatively characterize chaotic transport,  
 an important basic physical process occurring in many contexts, from fluid dynamics
\cite{aref-ottino,diego} to fusion plasma confinement \cite{plasmahorton}.  A now popular 
method for  determining chaotic transport  is based on {\it Lagrangian Coherent Structures} 
(LCS) \cite{haller}.  LCS represent effective barriers  that separate 
 regions of different dynamical behavior.  This FI has been used to investigate
time-dependent flows occurring in many applications of fluid mechanics,
such as  transport in ocean currents \cite{shay}, flow over an
airfoil \cite{haller04}, and chaotic mixing in forced tanks
\cite{gollub}, and it   has recently been introduced into plasma physics and used to describe 
plasmas turbulence \cite{jenko} and magnetic reconnection  \cite{grasso}.

In practice LCS are unveiled via computation of FTLEs for many points, 
i.e.\  the FTLE field. One then finds  ridges in this FTLE field that inhibit transport between different
regions. Such ridges, although  not {\it bona fide} invariant sets,  provide effective transport barriers.

% Even in time-independent systems LCS are important  for describing non-uniform transport. 

The detection of LCS by FTLEs is versatile -- it  can  be applied to periodic, 
quasiperiodic,  or  broadband vector fields defining the flow.  However, in two-dimensions when the vector field is  area-preserving and  contains a single temporal frequency, the FTRN proposed in this article is a better alternative.  Although for tractability many early studies of mixing in fluids \cite{aref-ottino} considered single-frequency vector fields,  physical  velocity fields contain many frequencies or  broadband turbulent spectra.  Moreover,   area preservation in fluids arises from the solenoidal  approximation of the velocity field, a common assumption, of e.g.\ geophysical  interest (e.g.\ \cite{diego,vera}),  with varying degrees of validity.  However,  for analyzing snapshots of magnetic fields in toroidal plasma devices or,  more generally,  astrophysical or other magnetic fields  in the vicinity of any stable closed field line  \cite{tokamaks}, the time-like coordinate is a  toroidal angle \cite{grasso} and the system contains  exactly only a single frequency.  Also, area preservation is an exact consequence of $\nabla \cdot {\bf B}=0$.  Thus for  such systems,   the FTRN  is a natural and  better suited fast indicator for characterizing chaos and transport.

For single frequency systems,  we show results on LCS using the  FTLE can be obtained by the  simpler and computationally faster method based on the FTRN.  As for the FTLE, LCS are ridges of the FTRN field computed from a grid of initial conditions.  But, the rotation number does not require the evaluation of spatial derivatives;  thus a fine mesh, although desirable, is not essential and this  substantially reduces  the computational time  to obtain LCS with good resolution. 

A time-$T$ periodic dynamical system with annular phase space $\mathcal{D}$ is determined by  a Poincar\'e map,   $M\colon\mathcal{D} \rightarrow  \mathcal{D}$,   where each period-T is represented by one iteration of $M$.  If  $\mathbf{x}\in \mathcal{D}$ lies on an invariant circle  ${S}^1$, then $M$  maps   ${S}^1$ to itself.  The rotation number (e.g.\ \cite{greene}) for  an orbit starting at  $\mathbf{x}_0$ is  $\omega = \lim_{n\rightarrow\infty} \Pi\cdot({M^n(\mathbf{x}_0) - \mathbf{x}_0})/{n}$  which is lifted to $\R$ and $\Pi$ is a suitable angular projection. Under mild conditions on $M$,   this limit exists for every initial condition $\mathbf{x}_0 \in {S}^1$ and does not depend on $\mathbf{x}_0$.  Consider a simple example, the rigid rotation $M(\mathbf{x}) = (x + w,y)$,  where $\mathbf{x}=(x,y)$ and $x\in{S}^1$.  Here the rotation number   $\omega = w$.   If $w$ is a rational number $p/q$, the trajectory is a period-$q$ orbit of the map $M$, whereas if $w$ is irrational, then the ensuing (quasiperiodic) orbit covers densely the circle ${S}^1$. The FTRN is the time-$NT$ truncation for the flow and corresponds in  the preceding definition to $N$ iterations of $M$.  Below we  use   $\omega_N(\mathbf{x}_0):=  \Pi\cdot (M^N(\mathbf{x}_0) - \mathbf{x}_0)/N$, with only $N\in\Z$ iterations.  In general, $\omega_N$, like any truncation,  depends on the initial condition.   While the infinite-time rotation number is not defined for chaotic orbits, which do not lie on any $S^1$,  the  finite-time counterpart exists for any orbit. Roughly speaking, $\omega_N$ measures the average rotation angle  swept out by a trajectory over a time interval $NT$, and thus conveys information about the local behavior of trajectories, just as the  FTLE does (which measures local rates of contraction or expansion).   We identify LCS with  ridges of the FTRN.  Note, minor ridges of the FTRN are finite pieces of invariant tori, which appear because the full rotation number has not been calculated.  

In order to illustrate how the  FTRN  detects LCS,  we consider three  examples and compare each with FTLE calculations.   Essentially the same results are  obtained with FTRN, but  with lower computational cost. One example is a time-periodic two-dimensional fluid flow,  the second is a discrete-time map of a flow used for passive advection, and the third is a  magnetic field-line map. In all cases  a uniform grid of $800\times 800$ points is advanced.

\smallskip

%%%%%%%%%%%%%%%%%%%%%%%%%

\noindent{\it 1. Periodic  double gyre flow.} We consider a two-dimensional fluid flow with a stream function,
\beq
\label{stream}
\psi(x,y) = A \sin [ \pi f(x,t)] \sin(\pi y),
\eeq
where $f(x,t) = a(t) x^2 + b(t) x$,   $a(t) = \epsilon \sin(2\pi t/T)$,  $b(t) = 1 - 2\epsilon \sin(2\pi t/T)$,   ${\cal D} := \{ 0 \le x \le 2, 0 \le y \le 1\}$, and   $A$ is the maximum horizontal  velocity, $u$ \cite{web}. The velocity field ${\bf v} = (u,v)$ is given by $u = -  \partial\psi/\partial y, \ v =  \partial\psi/\partial x$.

For $\epsilon=0$  the flow is integrable, with equilibrium points  $A:(1/2,1/2)$, $B:(3/2,1/2)$, $C_i:(x_{ci},y_{ci})$, $i=1,\ldots 6$, where $x_{ci} \in \{0,1,2\}$ and $y_{ci} \in \{ 0,1\}$. The points $A$ and $B$ are centers, whereas $C_i$ are saddles connected by heteroclinic trajectories. The latter are  boundaries of two gyres surrounding  $A$ and $B$, with clockwise and counterclockwise rotations, respectively. The heteroclinic trajectory ${\cal H}$ connecting $C_3:(0,1)$ and $C_4:(1,1)$ separates two distinct gyres and thus is a natural  place to focus  attention when looking for LCS.

For $\epsilon \ne 0$ the flow is  time-dependent and nonintegrable,  yet  ${\cal D}$ remains invariant. As is well-known, the former heteroclinic connections are structurally unstable; upon perturbation an  entanglement of stable and unstable manifolds with concomitant horseshoe dynamics appears. The vertical line ${\cal H}$ is no longer invariant,  but can be thought of as roughly separating two gyres with time-varying amplitudes: ${\cal H}$ oscillates in the horizontal direction with amplitude $\approx \epsilon$ (for small $\epsilon$) and frequency $2\pi/T$.  The   $T$-periodicity  permits  a  time-$4T$ rotation number,    $\omega_4(\mathbf{x}) =\sum^4_{i=1}\theta(\mathbf{x}_i)/4$, where $\tan \theta(\mathbf{x}) =  (x_{A,B}-x)/(y_{A,B}-y)$ is a rotation angle around gyre $A$ or $B$, for the time-$4T$ stroboscopic map of this nonintegrable system. This defines the projection $\Pi$.  Similarly,  we can compute the corresponding time-$4T$ Lyapunov exponent, which requires computing five nearby orbits,  necessary to evaluate spatial derivatives, instead of only one orbit for the FTRN.  For this reason, computation of FTRNs is at least five times faster than FTLEs. 

\begin{figure}
\includegraphics[width=1.0\columnwidth,clip]{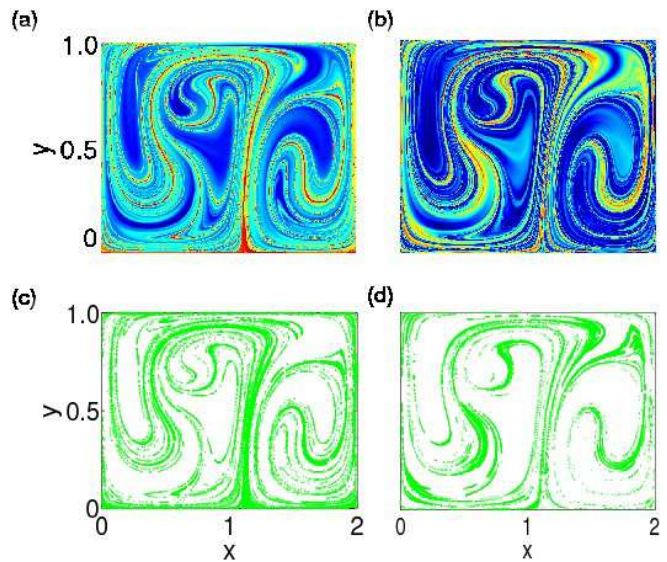}
\caption{\label{gyrefig} (color online) (a) Time-$4T$ Lyapunov exponent and (b) time-$4T$ rotation number for the double gyre system, with period $T=10$, amplitude $A=0.1$ and forcing strength $\epsilon=0.25$. (c) and (d) depict the Lagrangian Coherent Structures  corresponding to ridges of (a) and (b), respectively.}
\end{figure}

The results of the FTLEs and the FTRNs for the double gyre system are depicted in Figs.~\ref{gyrefig}(a) and (b), respectively, where blue (red) depict low (high) values of the corresponding quantity. The ridges of both, i.e.\  the crests of higher values are shown in Figs.~\ref{gyrefig}(c) and (d)  for FTLE and FTRN, respectively. The pictures are indeed very similar, notwithstanding the wide difference in the CPU-time necessary to produce them.  Moreover, Figs.~\ref{gyrefig}(c) and (d) reveal the existence of LCS for  the nonintegrable system. For small values of $y$,  the LCS approach the oscillating vertical line ${\cal H}$ that separates the gyres. In fact, despite the absence of well-defined stable and unstable manifolds of equilibria for time-periodic flows, the LCS displayed by Fig.~\ref{gyrefig} are quasi-invariant: if a passive scalar (tracer) were put on such LCS, it would be advected by flow and  remain in the vicinity of the LCS for a long time (on the order of the experiment duration). In other words, even though the LCS no longer separate the gyres for an arbitrarily long time (there may be a small transverse flux), trajectories starting on the lefthandside (righthandside) chiefly remain in the lefthandside (righthandside). In practical terms, however, this suffices to characterize an effective transport barrier.

\begin{figure}
\includegraphics[width=1.0\columnwidth,clip]{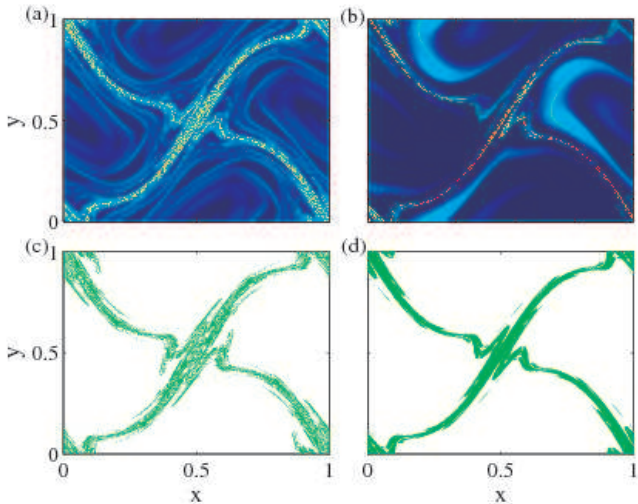}
\caption{\label{sinemap} (color online) (a) Time-$100T$ Lyapunov exponent and (b) time-$100T$ rotation number for the advection map (\ref{M}) with mixing parameter $a = 0.25$. (c) and (d) depict the Lagrangian Coherent Structures corresponding to ridges of (a) and (b), respectively.}
\end{figure}

%For our other two examples,  results obtained from FTRNs are so similar to those obtained from FTLEs that we shall only  display the former.
 
 \smallskip
 
 %%%%%%%%%%%%%%%%%%
\noindent{\it 2. Advection of a passive scalar.} In Example 1.,   investigation of LCS in a time-$T$ periodic flow  required the  numerical evaluation of the time-$T$ stroboscopic map $M$.  Sometimes when  researchers consider advection of a  passive scalar by  two-dimensional flows,  $M$ is given  by replacing the mixing action of a flow by an explicit area-preserving map.  An example   \cite{pierrehumbert}  of this is  
\bea
x_{n+1} & = & x_n + a \sin(2\pi y_n)  \quad (\mbox{\rm mod} 1), 
\nonumber\\
y_{n+1} & = & y_n + a \sin(2\pi x_n)  \quad (\mbox{\rm mod} 1)\,.
\label{M}
\eea
The map (\ref{M}) corresponds to a velocity field that is the  superposition of two sinusoidal shear flows in the $x$ and $y$ directions. In this case, the flow shear reverses sign along some shearless curve, as is the case for zonal flows of geophysical,  atmospheric, and plasma physical interest \cite{diego, tokamaks}. The map (\ref{M}) is symplectic and represents a Hamiltonian system for any  value of $a$.  The fixed points in the torus $\mathcal{D}=[0,1) \times [0,1)$ are the centers $P:(0,1/2)$, $Q:(1/2,0)$ and the saddles  $R:(0,0)$ and $S:(1/2,1/2)$. 

Since the system is nonintegrable for $a \ne 0$, the stable and unstable manifolds stemming from the saddle points $R$ and $S$ intersect in a heteroclinic tangle and there are chaotic orbits that do not line on continuous invariant circles. This structure is  responsible for the mixing effect of the chaotic advection.  Nevertheless,  this chaotic layer acts as an effective transport barrier separating the two gyres with invariant tori encircling $P$ and $Q$. 

This is clearly seen after computing the time-$100T$ rotation number for $M$ (with $\Pi$ projecting onto the $x$-axis) and then extracting the corresponding ridges with  high values of $\omega_{100}$. These ridges trace out quasi-invariant sets that shadow the heteroclinic connections, especially in the vicinity of the saddle points $R$ and $S$, reinforcing their interpretation as LCS.   The computation of FTRNs is particularly fast for maps, so calculation of $\omega_{100}$ is not difficult, but the results are essentially identical for $\omega_{10}$.

In Figs.~\ref{sinemap}(a) and (b) we depict the FTLE and FTRN, respectively, for orbits of the advection map (\ref{M}) for $a = 0.25$, whose ridges (points with largest relative values) are shown in Figs.~\ref{sinemap}(c) and (d), respectively. Both diagnostics indicate that the LCS are quasi-invariant sets about a  chaotic separatrix layer.  This layer acts as a transport barrier that separates  quasiperiodic curves encircling the centers $P$ and $Q$; the layer being a ridge implies  these are LCS of this system.

\smallskip

\noindent{\it 3. Magnetic field line map.} Magnetic field lines in the equilibrium states of toroidal magnetic plasma confinement devices, such as  tokamaks and stellerators,   are orbits of a one degree-of-freedom integrable Hamiltonian system, where a toroidal-like angle plays the role of time (e.g.\ \cite{tokamaks}).  In the simplest case,  canonically conjugate variables are the  spatial  coordinates of an annulus and points $(x_n,y_n)$ represent the  $n$-th field line intersection with a surface-of-section at a fixed value of the toroidal angle.  A similar situation arises in the vicinity of any stable (elliptic) closed magnetic field line. 

Perturbations due to external electric currents or internal MHD instabilities break symmetry and render the system nonintegrable, giving rise to chaotic field lines. Since the magnetic field configuration here is strictly static in time, the term chaos means that  two infinitesimally close field lines Lyapunov exponentiate as they  wind around the  torus. 

An example of a nonintegrable system that models such field lines is provided by the so-called {\it tokamap} \cite{balescu}:
\begin{eqnarray}
y_{n+1} & = & y_n- \frac{L}{2\pi} \, \frac{y_{n+1}\sin(2\pi x_n)}{1+y_{n+1}},\nonumber \\
x_{n+1} & = & x_n+\frac{1}{q(y_{n+1})}-\frac{L}{2\pi} \, \frac{\cos(2\pi x_n)}{(1+y_{n+1})^2}, 
\label{B}
\end{eqnarray}
where $L$ is a parameter measuring nonintegrability (proportional to the perturbation strength) and $1/q = (2-y)(2-2y+y^2)/4$ is the inverse of the rotational transform of the field lines. The value of $y_{n+1}$ is obtained by  applying Newton's method to (\ref{B}) at each map iteration.

\begin{figure}
\includegraphics[width=1.0\columnwidth,clip]{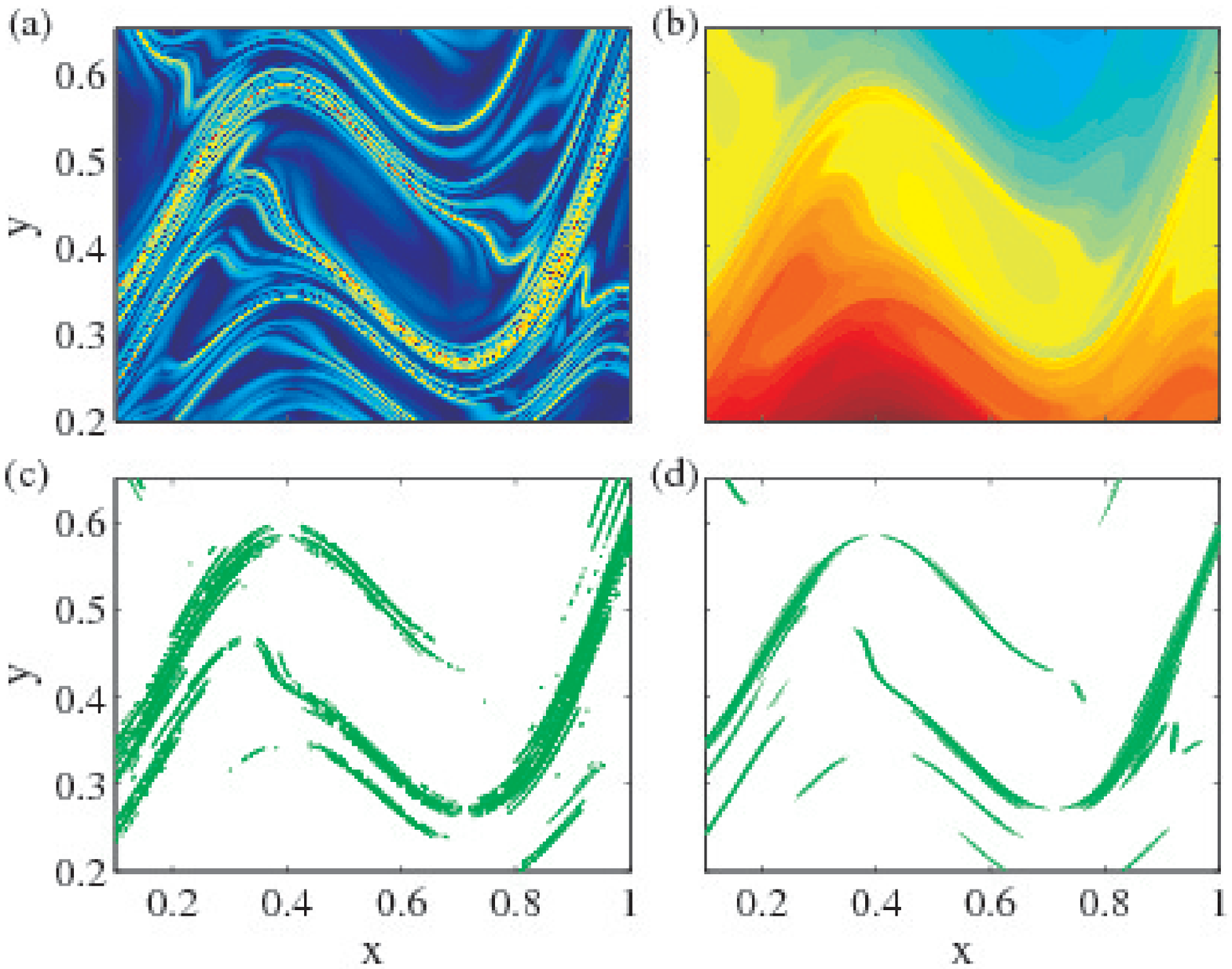}
\caption{\label{tokamapfig} (color online) (a) Time-$15T$ Lyapunov exponent and (b) time-$15T$ rotation number for the tokamap (\ref{B}) with $L=4$. (c) and (d) depict the Lagrangian Coherent Structures  corresponding to ridges of (a) and (b), respectively.}
\end{figure}

For $L = 3/2\pi$ the FTLE and FTRN (with $\Pi$ again the $x$-projection) of the tokamap are shown in Fig.~\ref{tokamapfig}(a) and (b), respectively, with (c) and (d) depicting  the corresponding ridges. In both cases the latter correspond, as in the advection map, to the homoclinic intersection of manifolds of the saddle fixed points of the tokamap (\ref{B}), defining a thin chaotic layer originating from such intersections. The separatrix layer is again a LCS, which indicates a transport channel along the layer,  yet  restricting diffusion across the  barrier. 

%%%%%%%%%%%%%%%%

In conclusion, here  we have proposed the FTRN as a FI, which for physical systems with a single period is superior, being faster and simpler, to the FTLE.  The three examples treated demonstrate this point.  We note, however, that the speed of calculation of both FTRNs and FTLEs is achieved at the cost of sacrificing  the detailed picture of transport provided  by an analysis of turnstiles through cantori  (e.g.\ \cite{mmp}), which requires a search for periodic orbits \cite{greene}.  In future studies we propose relaxing  the single frequency limitation  by a more detailed frequency analyses (e.g.\ \cite{laskar}), and considering  statistical analyses of the FTRN akin to that of  FTLE (e.g. \cite{frouard}).

%%%%%%%% 

This work was funded by  FAPESP, CNPq, CAPES, MCT/CNEN, Funda\cao Arauc\'aria, and the  USDOE Contract  DE-FG05-80ET-53088. The authors would like to thank D. Borgogno, D. Grasso, and T. Schep for stimulating discussions.

\end{document}